\documentclass[aps,prl,10pt,twocolumn,superscriptaddress]{revtex4-1}

\usepackage[utf8]{inputenc} 
\usepackage[T1]{fontenc}
\usepackage{amsmath}
\usepackage{amssymb}
\usepackage{gensymb}
\usepackage{bbm}
\usepackage{graphicx}
\usepackage{framed}
\usepackage{color}
\usepackage{hyperref}
\usepackage{epstopdf}
\usepackage{dsfont}
\usepackage{amssymb}
\usepackage{amsmath}
\usepackage{amsthm}
\usepackage{wrapfig}
\usepackage{bm}
\usepackage{enumitem}
\usepackage{mathtools}

\DeclareUnicodeCharacter{00A0}{~}

\usepackage{pifont}

\newcommand{\ba}{\begin{eqnarray}}
\newcommand{\ea}{\end{eqnarray}}

\begin{document}

\title{Measuring the time-frequency properties of photon pairs: a short review.}

\author{Ilaria Gianani}
\affiliation{Dipartimento di Fisica, Sapienza Universit\'a di Roma, Piazzale Aldo Moro, 4 - 00185 Rome, Italy}
\affiliation{Dipartimento di Scienze, Universit\'a degli Studi Roma Tre, Via della Vasca Navale, 84 - 00146 Rome, Italy}

\author{Marco Sbroscia}
\affiliation{Dipartimento di Scienze, Universit\'a degli Studi Roma Tre, Via della Vasca Navale, 84 - 00146 Rome, Italy}

\author{Marco Barbieri}
\affiliation{Dipartimento di Scienze, Universit\'a degli Studi Roma Tre, Via della Vasca Navale, 84 - 00146 Rome, Italy}

\begin{abstract}
Encoding information in the time-frequency domain is demonstrating its potential for quantum information processing. It offers a novel scheme for communications with large alphabets, computing with large quantum systems, and new approaches to metrology. It is then crucial to secure full control on the generation of time-frequency quantum states and their properties. Here, we present an overview of the theoretical background and the technical aspects related to the characterization of time-frequency properties of two-photon states. We provide a detailed account of the methodologies which have been implemented for measuring frequency correlations and for the retrieval of the full spectral wavefunction. This effort has benefited enormously from the adaptation of classical metrology schemes to the needs of operating at the single-photon level.
\end{abstract}

\maketitle 

\section{Introduction}

Increased capacity of communication systems brings in the demand for security standards able to keep up wit technological processes. This has prompted investigation on quantum light as means to distribute information \cite{reviewobrien, thew07}. Quantum-enabled security stems from the inextricable connection between the logical processing and physical substrate. Protection does not rest on computational complexity, but in the very functioning mechanism of the devices \cite{reviewgisin, bb, ekert91}. This perspective requires unprecedented level of control on light in order to make protocols robust and trustworthy\cite{Scarani2009, elenicomm}. 

Transposition of these ideas to actual implementations requires identifying which physical properties are more suitable to specific tasks. Transmission in the atmosphere may privilege light polarization, as this can be preserved~\cite{Decoy2007, Fedrizzi2009, Capraro2012, Vallone2015,
Liao2017, Yin2017, Shanghai2017}. On the other hand this degree of freedom is less suitable for fiber transmission, for which internal degrees of freedom are a far better option; these include quadrature of light \cite{grangiernat,eleniphot} and the frequency-time domain \cite{gisin00,franson}.

Nowadays there exist a solid effort in developing the time-frequency platform into full technological maturity, inspired by the potential this unleashed in classical communication. The first application of these degrees of freedom has been the use of entangled time bins for quantum communications \cite{gisin03,cabello09,gisin04,gianani11,cuevas13,tittel16,ursin17,vedovato18,shields18}. In this scheme each photon of a pair occupies a time bin corresponding to either an early or a late arrival time, with respect to a clock. Since the two photons are emitted at the same time, i.e. they are correlated in the time-frequency domain, their state can be prepared as a quantum superposition of them both arriving early or late. Since their introduction in Ref.~\cite{franson}, time bins have been established as an efficient solution for distributing quantum information over long distance in fibre. Encoding in frequency bins has also been approached as a viable alternative, due to their intrinsic resilience to spectral dispersion, although direct manipulation of frequencies at the single photon level may be demanding \cite{nathan1,massar1,kobayashi,morandotti,optica}.

More recently the control of the spectral-temporal envelope down to the single-photon level has reached such a level of sophistication that it is conceivable to encode quantum information in different time-frequency modes \cite{uren06,benni112,benni11,vahid18}. It has been demonstrated that these offer a complete platform for quantum information processing \cite{benni15}. Experimental progress has been steady, delivering convincing prototypes of elementary gates. This has been further expanded designing quantum networks linking distinct time-frequency modes with quantum correlations in their quadratures: this constitutes an excellent platform for implementing continuous variables one-way quantum computing. The time-bin and continuous variable schemes have been merged to implement very large cluster states; different modes corresponding to distinct time (or frequency) bins are manipulated by means of quantum interferometry, so that they share quadrature entanglement~\cite{Menicucci2008, Roslund2013, Chen2014, Yokoyama2013, Asavanant2019, Larsen2019}. 
The basic physical process that has enabled the development of these new technologies is time-frequency entanglement, naturally present in spontaneous parametric down conversion (SPDC) \cite{spdc,spdc1,spdc2,spdc3,spdc4,spdc5,Karpinski2009}, as well as spontaneous four-wave mixing (FWM) \cite{kumar01,rarity07,brian09,bajoni1,bajoni2}.  Assuring its presence in a reliable way is by no means a simple task since it demands addressing quantum light at ultrafast timescales. 

The last three decades have seen the development of lasers capable of delivering pulses with increasingly larger bandwidths. Their use and characterization has required a great effort for the identification of the correct indirect measurement to extract information on the temporal profile, as the achieved timescales were shorter than the response of the electronic equipment used for direct measurements (e.g. photo diodes) \cite{Walmsley2009}. This has been even more the case when more involved sources as high harmonics \cite{corkum} or coherently synthesised pulses \cite{goulielmakis,manzoni} have been investigated, with promising outlooks for their use in applied and fundamental physics alike.
Due to the prohibitive time scales, in order to retrieve the temporal structure of the generated pulses, indirect measurements acquiring the spectral amplitude and phase have been proposed.

In this review, we describe how ideas from classical ultrafast metrology have influenced positively new techniques and methods for characterization of time-frequency correlations of photon pairs emitted by SPDC or FWM. We provide the necessary background on both subjects. As quantum communications are fostered towards higher technology levels, a multi-disciplinary spectrum of competences will be required to active specialists. We provide here a hopefully useful guide.

\begin{figure*}[t!]
    \includegraphics[width=0.30\textwidth]{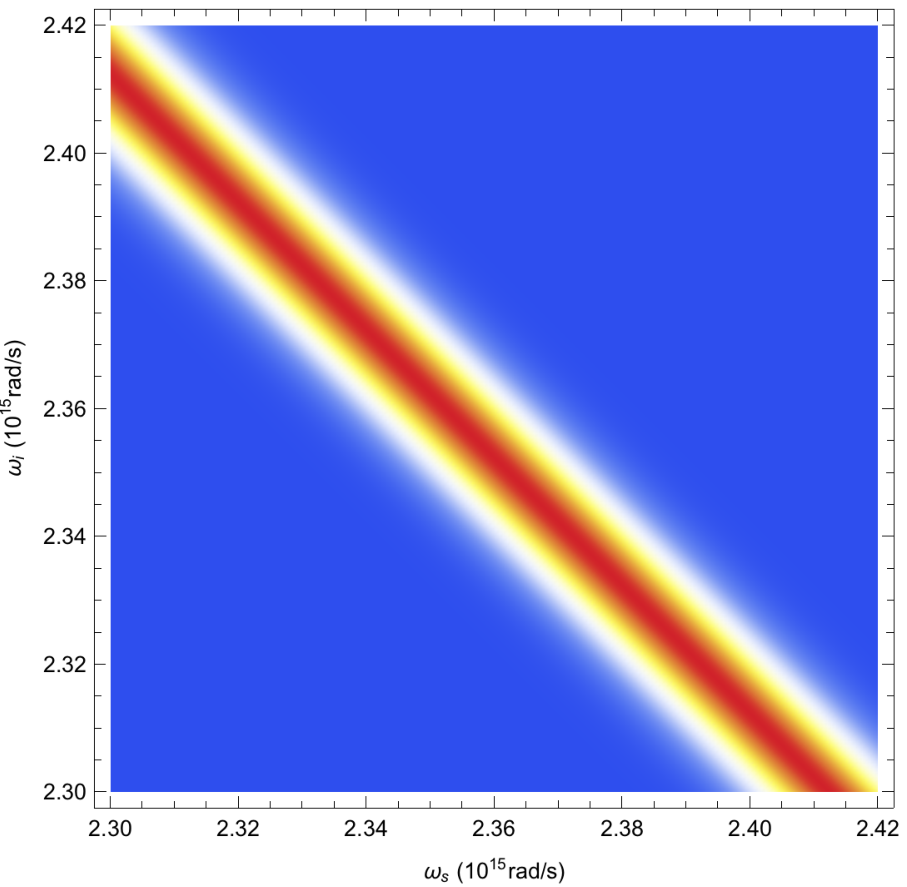}
    \includegraphics[width=0.30\textwidth]{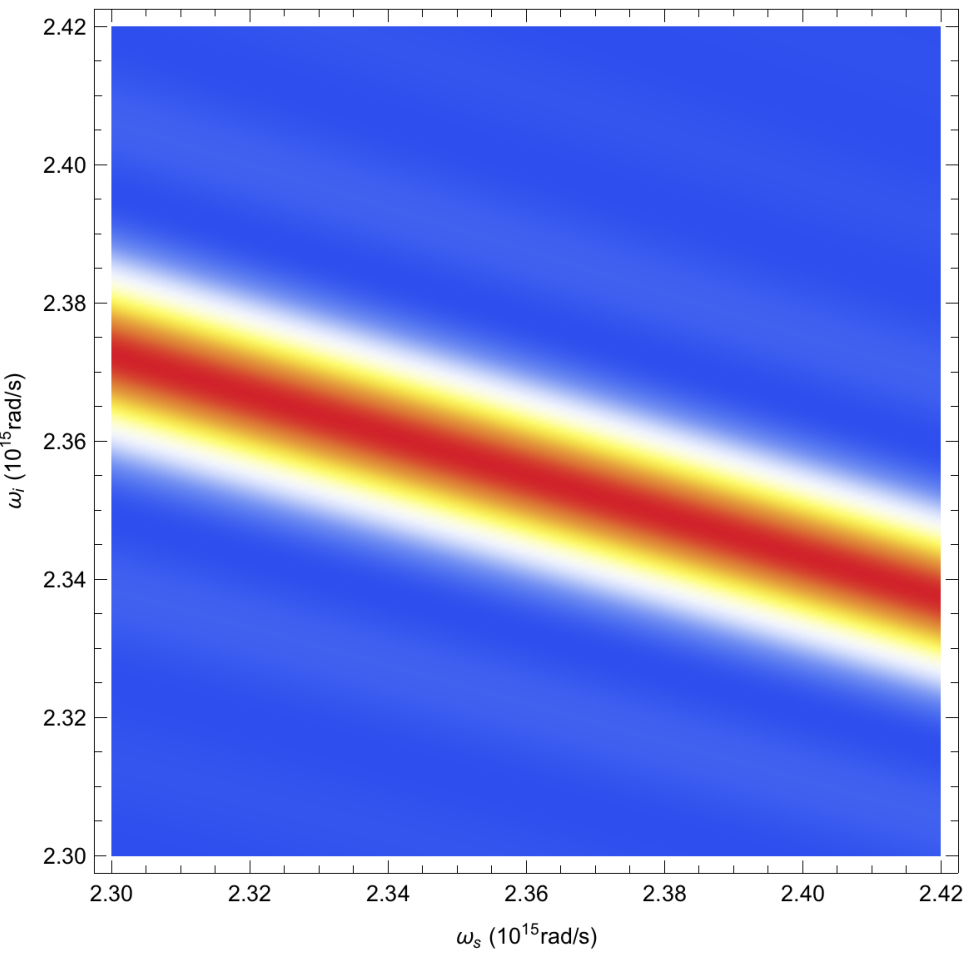}
    \includegraphics[width=0.30\textwidth]{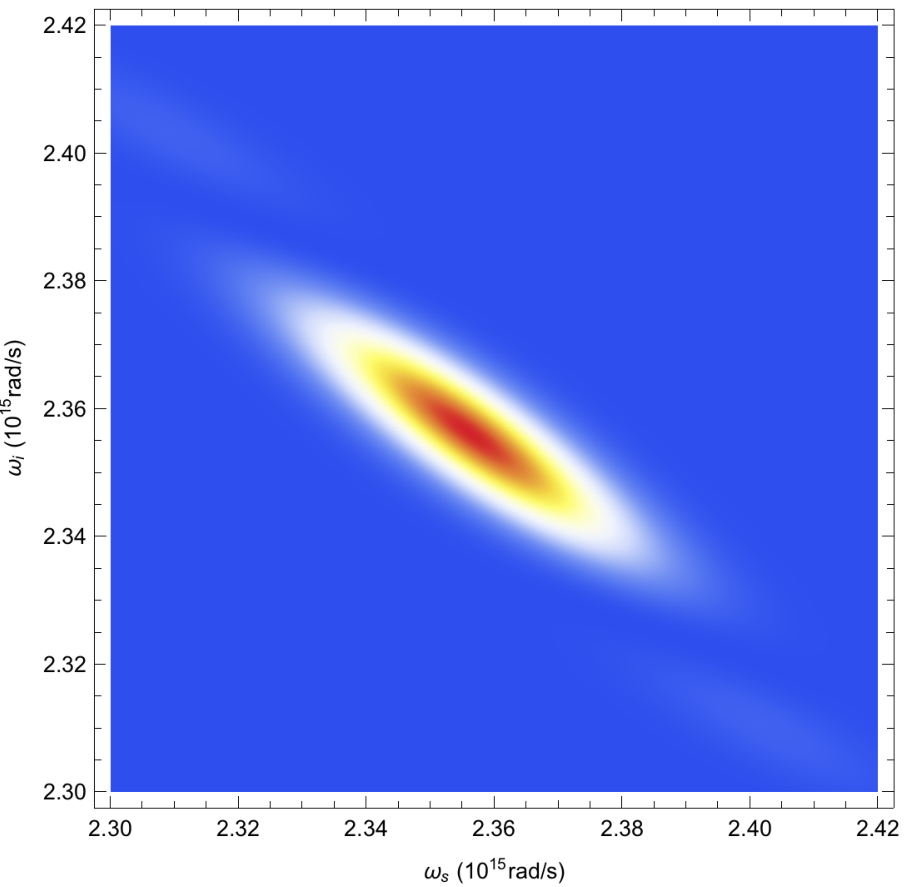}
    \caption{Joint spectral intensity for a two-photon pair. a) Phase matching function $|\phi(\omega_s,\omega_i)|^2$ for a 1-mm BB) crystal, cut for degenerate, type-II phase matching ($\lambda_p=800$nm). b) Spectral intensity of the pump beam with a Gaussian profile of full width at half maximum $\Delta \lambda=1$nm. c) resulting JSI.}
    \label{fig:JSIasual}
\end{figure*}
\section{the time-frequency framework}

We sketch the theoretical treatment for a free-space source of correlated photons based on parametric down-conversion. Adoption of waveguided solutions can be obtained in a similar way, by taking into account proper dispersion relations; the same applies to sources based on four-wave mixing. The pump beam has a spectral profile in the form:
\begin{equation}
\label{pompa}
\alpha(\omega_p)=|\alpha(\omega_p)|e^{-i\varphi(\omega_p)}.
\end{equation}
The paraxial approximation is taken, hence we neglect the transverse profile - it should be noticed, however, that space-time coupling might arise if tight focusing is adopted. 

The dispersion relation in the crystal associates to each frequency in the pump envelope~\eqref{pompa} a momentum $k_p(\omega_p)$. The parametric down-conversion process delivers a pair of photons, whose energies and momenta should obey conservation laws:
\begin{equation}
\label{perkappa}
        k_p(\omega_p)=k_s(\omega_s)+k_i(\omega_i),
        \end{equation}
        and
        \begin{equation}
        \label{peromega}
    \omega_p=\omega_s+\omega_i.
\end{equation}
The labels for the two photons reflect the customary choice of the names signal and idler to identify them, with the idler being the one at lower energy \footnote{This terminology is borrowed from microwave technology, in which parametric processes are used for low-noise amplification. The analogy with non-linear optical case rests in the fact that energy storing quantities are modulated at a frequency $\omega_p$ (the electronic polarization in a nonlinear crystal, or the capacitance in the circuit).
In the typical arrangement, signal and pump are at close frequencies, thus generating a third frequency in the radio range.}. These names are also kept in the frequency-degenerate case. As for the transverse spatial profile, it is commonplace to select a single spatial mode for each photon by means of optical fibres.

While the energy conservation~\eqref{peromega} can be enforced due to the long duration of the nonlinear interaction, the momentum conservation is influenced by the finite longitudinal size $L$ of the crystal; while the perfect case would give a Dirac delta in the momentum wavefunction $\delta(k_p(\omega_s+\omega_i)-k_s(\omega_s)-k_i(\omega_i))$, this is regularised as
\begin{equation}
\Phi(\omega_s,\omega_i) =\text {sinc}\frac{\Delta k(\omega_s,\omega_i) L}{2},    
\end{equation}
where $\Delta k(\omega_s,\omega_i)=k_p(\omega_s+\omega_i)-k_s(\omega_s)-k_i(\omega_i)$. The two-photon wavefunction is then written as: 
\begin{equation}
    |\Psi\rangle=\int d\omega_sd\omega_i\, \alpha(\omega_s+\omega_i)\Phi(\omega_s,\omega_i)\,\hat a^\dag(\omega_s)\hat a^\dag(\omega_i)|0\rangle.
\end{equation}
Here $\hat a^\dag(\omega)$ denotes the creation operator on a mode with frequency $\omega$; the photons are thus made distinguishable by either a polarisation or a spatial label, as commonplace in the experiments.
\begin{figure}[b]
    \centering
    \includegraphics[width = \columnwidth]{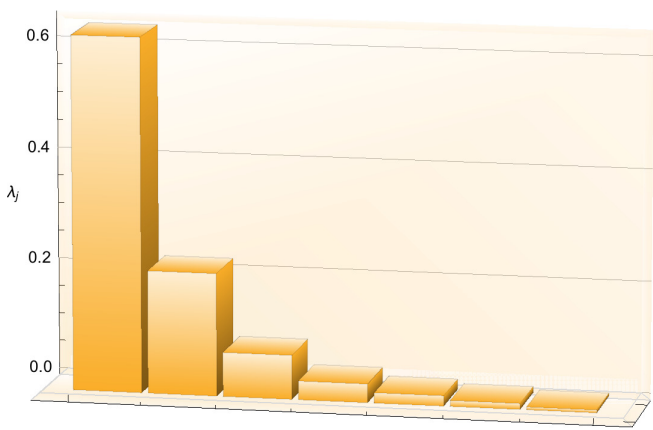}
    \caption{Weights of the Schmidt decomposition of the JSA.}
    \label{fig:lambdas}
\end{figure}{}
Frequency correlation properties of the photons are contained in the joint spectral amplitude (JSA)\cite{Cassemiro2010a,Grice1997,Grice2001}:
\begin{equation}
\label{eq:JSA}
    J_A(\omega_s,\omega_i)=\alpha(\omega_s+\omega_i)\Phi(\omega_s,\omega_i),
\end{equation}
which contains information on both the pump spectral profile and the nonlinear optical response. Inferring information on $J_A(\omega_s,\omega_i)$ is the ultimate goal of an experiment aiming at characterising the spectral-time properties of the photon pair. It must be noticed that $J_A(\omega_s,\omega_i)$ is a complex function, taking into account the spectral amplitude as well as the spectral phase, while the energy distribution is captured by the joint spectral intensity (JSI) $J_I(\omega_s,\omega_i)=|J_A(\omega_s,\omega_i)|^2$; an example is reported in Fig.~\ref{fig:JSIasual}.

It has been realised that, while the support of $J_A(\omega_s,\omega_i)$ is continuous, it admits a decomposition over a discrete set of Schmidt modes \cite{Law2000,Eberly2006}
\begin{equation}
   J_A(\omega_s,\omega_i)=\sum_j \sqrt{\lambda_j} \psi^{s}_j(\omega_s)\psi^{i}_j(\omega_i), 
   \label{decomp}
\end{equation}
with only the first few modes actually giving a sizeable contribution under ordinary conditions; Fig.~\ref{fig:lambdas} shows this effect for the parameters of Fig.~\ref{fig:JSIasual}. Each pair of modes $\psi^{s}_j(\omega_s)\psi^{i}_j(\omega_i)$ then represents an effective source, independent on the others, by which the photons can be emitted with probability $\lambda_j$. The Schmidt number
\begin{equation}
    K=\frac{1}{\sum_j \lambda_j^2}
\end{equation}
provides a quantitative estimation of the number of modes involved in the emission process. In the limit $K=1$ the source generates spectrally pure single photons. 

The Schmidt number can be extracted by a measurement of the second order correlation function $g^{(2)}(0)$ of one of the modes. Indeed, the decomposition in Eq. \eqref{decomp} implies that each independent mode produces thermal light. The resulting multi-thermal emission has\cite{Christ2011} $g^{(2)}(0)=1+1/K$.

\section{Ultrafast metrology}

Complete measurements of the JSA (Eq. \eqref{eq:JSA}) demands assessing both its amplitude ad its phase. While the amplitude is easily accessed, most difficulties lie in the phase retrieval. 
This poses the same order of problems as the metrology of ultrashort pulses. This analogy constitutes a powerful resource since it allows for transfer of knowledge from the classical to the quantum domain, although differences in the technical and conceptual aspects remain ad should be addressed.
We review some popular techniques giving access to spectral phase in classical metrology to provide context for applications to the JSA reconstruction. 

The path to the optimal indirect phase measurement has been twofold: on one hand interferometric techniques have been exploited, providing direct access to spectral phase; on the other hand, spectrography has allowed to retrieve intensity traces from which phase information can be retrieved. Both approaches have been declined into various techniques to account for the most diverse scenarios, and offer up and downsides depending on the particular sought implementation.

Spectrographic techniques have been among the first attempts to characterize optical pulses. The concept behind these range of techniques is to retrieve the information about the pulse which is captured as a joint representation of the Fourier conjugate variables time and frequency. 
In a spectrogram the signal's spectrum is recorded as a function of the time delay with respect to the pulse peak. Mathematically a spectrogram is defined by the function:

\begin{equation}
    S(\omega,\tau)=\left\vert\int^\infty_{-\infty}E(t)g(t-\tau)e^{-i\omega t}dt\right\vert^2
    \label{spectrogram}
\end{equation}
where E(t) is the electric field and $g (t - \tau)$ is known as the gating function which selects sections of E(t) as the delay $\tau$ varies. The probe and gate functions can be either obtained from the same identical pulse, in which case the method is self-referencing, or from different pulses in which case a known reference can be used.
Although the spectrogram is an intensity measurement and thus does not provide any direct information on the phase, it is enough in most cases to fully characterize the field $E(t)$, however this makes the method prone to ambiguities in the reconstruction. These can be removed by constraining the algorithm through additional measurements.

\begin{figure}

\includegraphics[width= \columnwidth]{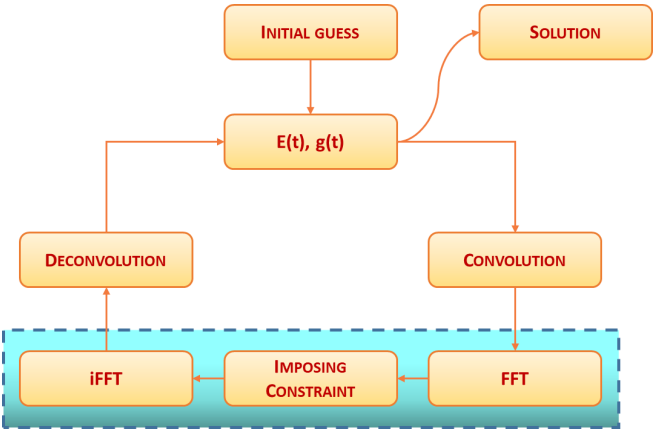}
\caption{FROG reconstruction algorithm.}
\label{schemafrog}
\end{figure}

The most common example of spectrographic technique is  Frequency Resolved Optical Gating (FROG) \cite{Trebino1993,Trebino1997, Trebino2000}. FROG is an autocorrelation-like measurement in which the upconverted signal from an intensity autocorrelator is spectrally resolved and recorded as a function of the time delay between the test pulse and a gate pulse. This results in a $N \times N$ points trace, where $N$ is the sampling size, from which $2N$ points (intensity and phase of the test pulse) are retrieved by means of an iterative algorithm. There are different types of FROG implementations which differ from the geometry and the gating function used, each tailored to specific measurement requirements \cite{Linden1998,DeLong1994, Trebino1997,Kane1993a,Kohler1995}.
FROG uses an iterative algorithm to extract the spectral phase from a spectrogram, the principal component generalised projections algorithm (PCGPA) \cite{Kane1997}. This algorithm is used to reconstruct the two fields that contribute to the FROG trace. Since this trace is a measured real valued quantity, the information it bears o the phase is not as direct as that in an interferogram, hence the need for an iterative reconstruction. In principle the two fields can be different, which is the case of blindFROG \cite{Kane1997}, but more ambiguities arise which are instead confined when the two pulses are identical, or the reference is known.

Interferometric techniques are based on spectral shearing interferometry (SSI) \cite{Wong1994}, where the phase information is stored in the fringe pattern given by the interference between a pulse and an identical time-delayed copy sheared in frequency. The most known example of SSI is Spectral Phase Interferometry for Direct Electric-field Reconstruction (SPIDER) \cite{Iaconis1998, Iaconis1999, Walmsley2009}, where the shear between the two time-delayed replicas is obtained by means of a nonlinear process. This is made possible by the use of a highly-chirped ancillary beam: the two replicas of the test pulse are focused on a nonlinear crystal where they are upconverted with a chirped ancillary pulse derived from the same original pulse. Since the two replicas are delayed in time by an amount $\tau$, they will upconvert with two different quasi-monochromatic frequencies from the chirped pulse. This will generate two signal pulses sheared in frequency by an amount $\Omega=\delta \tau$ where $\delta$ is the chirp rate.
The interference between the two signal pulses will read:

\begin{equation}
\begin{aligned}
&I_{SPIDER}(\omega) = \vert E(\omega)\vert^2 + \vert E(\omega-\Omega)\vert^2\\
&+ 2\vert E(\omega)\vert\vert E(\omega-\Omega)\vert \cos(\phi(\omega)-\phi(\omega-\Omega)+\omega\tau),
\end{aligned}
\label{interf}
\end{equation}

\begin{figure}

\includegraphics[width= 0.4\columnwidth]{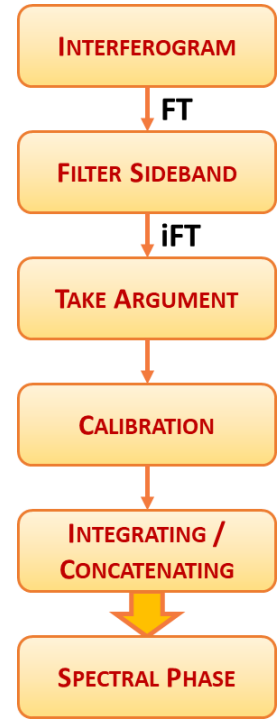}
\caption{SPIDER reconstruction algorithm.}
\label{figspider}
\end{figure}

Differently from spectrographic techniques, the spectral phase can be extracted directly from the interferogram. 
The oscillatory term containing the information on the phase, is constituted by a carrier $\omega\tau$ which results in spectral fringes with spacing of $2\pi/\tau$, modulated by the difference between the spectral phases of the two replicas. 
The phase information can be extracted via the reconstruction process described in Fig. \eqref{figspider}, based on Fourier filtering, known as the Takeda algorithm \cite{Takeda:82}. Since two sidebands are present in the Fourier transform of Eq. \eqref{interf}, if the delay $\tau$ is large enough, it is possible to isolate one of these by amplitude filtering. Then, performing the inverse Fourier Transform on the filtered sideband allows to extract the phase difference: $\,\Theta(\omega)=\phi(\omega)-\phi(\omega-\Omega) + \omega\tau,$ where the $\omega\tau$ term can be removed with a calibration step consisting in taking a measurement at $\Omega = 0$.
Retrieving the spectral phase $\phi(\omega)$  is then made possible either via integration or via concatenation of the phase difference $\Theta(\omega)$: once the phase for a given frequency $\omega$ is arbitrarily set, the phase on the next sampling point, $\omega+\Omega$ can be computed by $\phi(\omega+\Omega)=\phi(\omega)+\Theta(\omega+\Omega)$. This poses a limitation to SPIDER,and to all techniques reconstructing the envelope alike, as the phase will be retrieved up to a term linear in $\omega$.  \footnote{This reflects the ambiguity on the phase that the electric field acquires at the maximum of the envelope}

Since the development of SPIDER many different variants have been realized to improve and adapt the technique to respond to the different needs dictated by the pulses' complexity \cite{Dorrer2001a, Baum2004,Mairesse2005, Radunsky2007, Anderson2008,Austin2010b,Witting2009a,Kosik2005,Anderson2008b}; these included a referenced implementation, X-SPIDER \cite{Londero2003, Morita2002,Hirasawa2002}, where a known reference is used to make up for low intensity pulse, and SEA-CAR-SPIDER \cite{Witting2009}, which adopts a multiple shear approach to account for highly modulated spectra and improving the robustness to noise. 

The multi-shear approach has been further exploited in another interferometric technique stemmed from SPIDER, Mutual Interferometric Characterization of Electric-fields (MICE) \cite{Bourassin-Bouchet2013}, which makes use of the redundancy of information granted by the multishear implementation for reconstructing more than one field at once. In MICE, two fields depending on a set of parameters $\gamma$, $E_1(\gamma)$ and $E_2(\gamma)$, are made interfere after applying to $E_2(\gamma)$ a set of K shears $\Gamma_k$ with the same dimension as $\{\gamma\}$. 
The interferogram is processed with the Takeda algorithm, thus isolating one of the measured interference sidebands $AC_{meas}$. The contribution from the two different fields is then separated through a least square approach.Defining the error:
\begin{equation}
    \mathcal{E}=\sum_{j,k}=\left\vert AC^{meas}_{j,j-k} - E_1(\gamma_j)E_2^*(\gamma_j-\Gamma_k) \right\vert^2,
\end{equation}
it is possible to obtain the two fields by minimizing it with respect to both $E_1$ and $E_2$ imposing $\partial\mathcal{E}/\partial E_1 = 0$ and $\partial\mathcal{E}/\partial E_2 = 0$, which lead to the following equations:
 \begin{equation}
     \begin{aligned}
     &E_1(\gamma_j)=\frac{\sum_k AC^{meas}_{j,j-k}E_2(\gamma_j-\Gamma_k)}{\sum_k \left\vert E_2(\gamma_j-\Gamma_k)\right\vert^2}\\
     &\\
     &E_2(\gamma_j)^*=\frac{\sum_k AC^{meas}_{j,j+k}E_1^*(\gamma_j+\Gamma_k)}{\sum_k \left\vert E_1(\gamma_j+\Gamma_k)\right\vert^2}
     \end{aligned}
 \end{equation}
 
These equations are solved in an iterative algorithm starting with a random guess for one of the two fields, which is used to obtain a solution for the other field, until the error is minimized. In this way, each field acts as a reference for the other, hence constituting a mutually-referenced technique.

\section{Time-frequency correlations}

Whenever a single aspect present in the JSA is emphasised for the sought application, one can rely on methods tailored to capture that single feature. This is the case for instance of correlations in one degree of freedom. The limited access to information on the state is compensated by a resource-economic experimental scheme. 
\begin{figure}[b!]
\includegraphics[width=\columnwidth]{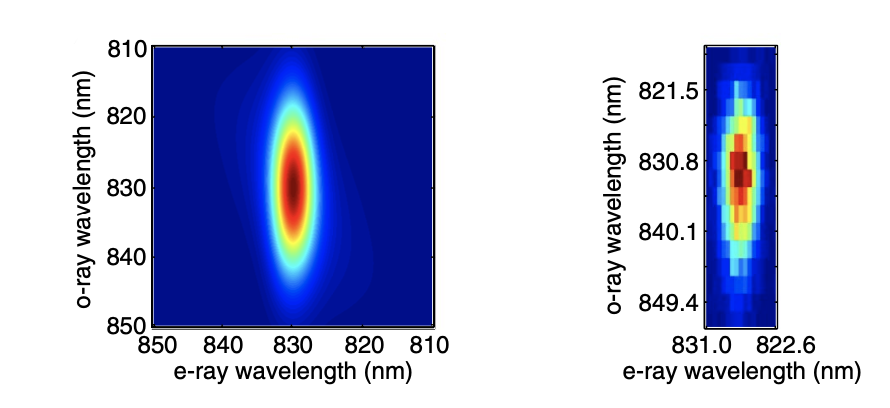}
\caption{{\it JSI reconstruction.} Expected and reconstructed JSI. Adapted from Ref.\cite{Mosley2008}.}
\label{figmosley}
\end{figure}
Frequency correlations are the easiest feature that can be characterized, since it only demands looking at correlations in the outcomes of two spectrometers. There exist two main strategies to implement these instruments at the single photon level: diffraction gratings can be employed to separate frequency components onto distinct spatial positions, which are then isolated by means of fibers. Alternatively, light can be coupled into a highly dispersive fibres that maps frequencies into different detection times \cite{xstine2009}.  Although this latter technique is limited by the time jitter of the detectors it has the advantage of requiring no raster scan based on a single detector. Both techniques are commonly adopted. As an example, in \cite{Mosley2008} a monochromator is employed to verify  the production of uncorrelated photons:  the two-photon Type II PDC from a potassium dihydrogen phosphate (KDP) crystal pumped at 415 nm with a 50 fs duration is collected and analyzed. The group velocities of the ordinary polarized photon and that of the pump are equal, allowing PDC to produce a frequency separable state. The expected and reconstructed JSI are reproduced in figure \ref{figmosley}, demonstrating good agreement. These measurements provide good indication on the factorability of the biphoton state, however, in the absence of phase information, this needs to be supported by an independent evaluation of the Schmidt number by a measurement of the second order correlation function \cite{HOMnoi}. 

However, it might happen that under narrowband pumping the width of the JSA in the two directions $\omega_s+\omega_i$ and $\omega_s-\omega_i$ covers very different ranges: while the difference can span several nanometers, the sum, governed by the pump spectrum, can be confined in the hundreds of MHz range. Encompassing both regimes with a single device would pose considerable constraints on the equipment, which are practically hard to meet. 
An interesting solution consists in coupling frequency to a different degree of freedom of the photons, foremost polarization, and inspect this instead. This strategy has been implemented in Ref.\cite{piilo} and in Ref. \cite{Sbroscia:18}, which we review here. 

When light travels through a dispersive birefringent material, the spectral phase accumulated differs for the two polarization components,on and off the fast axis. As a result, a photon initially in a pure state may emerge with a degree of mixing, depending on its polarization and on the birefringence dispersion. This effect can be quantified by means of polarimetric measurements: if the input polarization is at $45^{\circ}$ with respect to the material's axis, the contrast of a Malus measurement will drop. 

\begin{figure}
\includegraphics[width= \columnwidth]{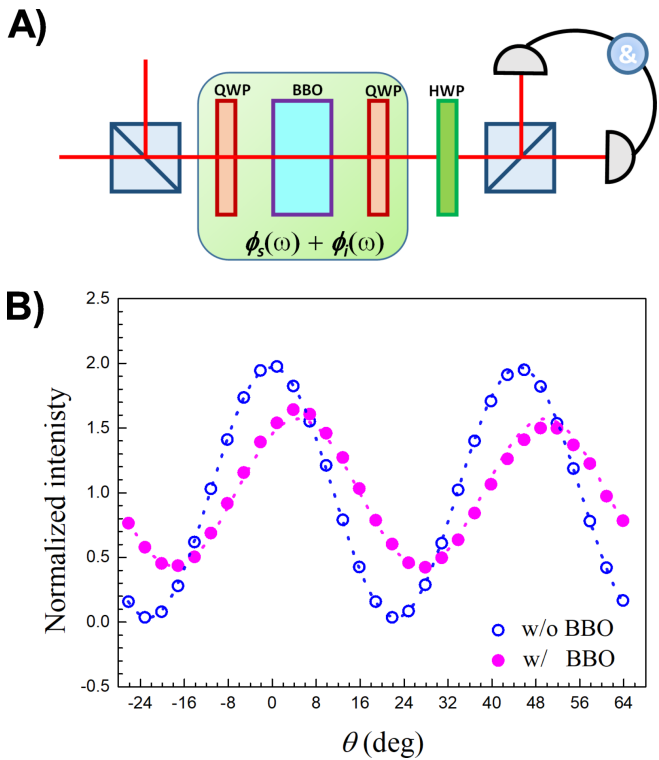}
\caption{{\it Frequency correlations.} a) Experimental setup: photons produced via type I - SPDC are superimposed on the same spatial mode resulting into a N=2 N00N state in the circular polarisation basis. The phase shift is applied through a combination of two quarter wave plates (QWP) and a BBO crystal. Coincidence count rate, as a function of the angle of the half wave plate (HWP) are collected. b) Coincidence rate as a function of the $\theta$ setting of the HWP with and without the BBO crystal. Data are normalised to their mean value. The figure has been adapted from Ref.\cite{Sbroscia:18}.}
\label{figol}
\end{figure}

\begin{figure*}[t!]
\includegraphics[width=0.8\textwidth]{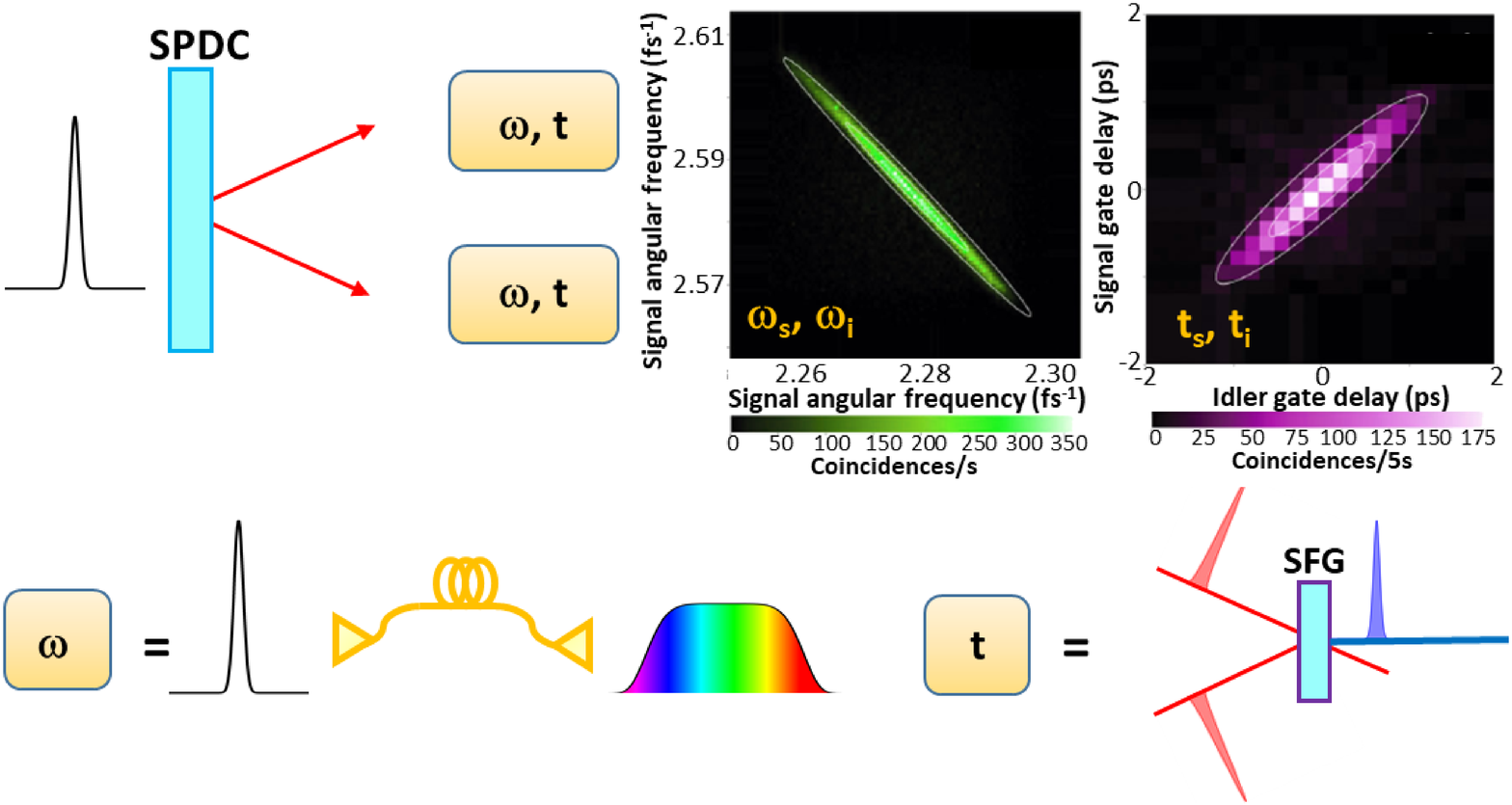}
\caption{{\it Direct entanglement quantification.} Experimental setup: spectrally broad photon pairs are produced via SPDC. Coincidence counts  recorded both in the temporal and frequency domain. The intensity profiles so obtained in the two domains are shown. The figure has been adapted from Ref.\cite{John2018} American Physical Society.}
\label{figjohn}
\end{figure*}
When using two-photon states, the latter effect can be reduced in the presence of frequency correlations. Consider the case of perfect negative correlations: any frequency $\omega_s$ will be correlated uniquely to a single $\omega_i=\omega_p-\omega_s$. The two photons are injected in the medium in the quantum state: 
\begin{equation}
\vert \psi(\omega_s,\omega_i)\rangle=\frac{1}{\sqrt{2}}\left(|f\rangle_s|f\rangle_i + e^{i(\phi_s(\omega_s)+\phi_i(\omega_i))}|s\rangle_s|s\rangle_i\right)    
\label{state}
\end{equation}
where $\vert f \rangle$ denotes the state polarized along the fast axis of the medium and  $\vert s \rangle$ along the orthogonal one. The phases acquired by the photons can be expanded as:
\begin{equation}
\begin{aligned}
&\phi_s(\omega_s)+\phi_i(\omega_i)= \phi_s(\omega_p/2)+\phi_i(\omega_p/2)+\\
&\phi'(\omega_p/2)\Delta\omega_s+\phi'(\omega_p/2)\Delta\omega_i,
\end{aligned}
\end{equation}
with $\Delta\omega_j=\omega_j-\omega_p/2$, and $\Delta\omega_s=-\Delta\omega_i$ due to perfect anticorrelations.  Hence, although the state \eqref{state} must be averaged on the frequencies, there will be  no chromatic effect on the phase, a phenomenon akin to dispersion cancellation []. Therefore, measuring the purity of the averaged output state would give, for arbitrary correlations, a measurement of the correlation strength. This can be quantified looking at the width of the JSA along the $\omega_s+\omega_i$ direction, that is reduced by a factor $\kappa$ with respect to the uncorrelated case; in turn, $\kappa $ can be estimated from the contrast of a two-photon Malus measurement. 

The experiment has been performed by superimposing two photons on the same spatial mode, in order to produce the state \eqref{state}. The photons are produced from SPDC by a 3mm $\beta$-barium borate (BBO) crystal cut for frequency degenerate Type I phasematching ($\lambda_p=405\,nm$, CW operated), and the dispersive phase is imparted by a similar crystal. The experimental setup is described in fig \ref{figol} (a). 
The collected data are the coincidence curves obtained with and without the insertion of the BBO. The reduced contrast is related to the width of the spectral phase distribution, hence on the reduction factor $k$. The experiment has delivered the value $\kappa=0.14\pm0.02$. We stress that this number cannot be directly related to the entanglement content of the JSA since it only quantifies the {\it classical} degree of correlation in the frequency domain.  

The ability to estimate the entanglement demands to access to two canonically conjugated degrees of freedom, which in our case are frequency and arrival time of the photons. 
Uncertainty relation can be used as a tool for discriminating entangled out of separable biphoton states. By labelling the quantum state of each photon using both the frequency, $\omega$, and the arrival time onto a detector, $t$, the product of correlations for separable states is lower bounded by
\begin{equation}
\Delta(\omega_s+\omega_i)\Delta(t_s-t_i)\geq1
\label{prodotto}
\end{equation}
where $\Delta(\omega_s+\omega_i)$ is the width of the joint spectral intensity, $J_I (\omega_s,\omega_i)$, and $\Delta(t_s-t_i)$ that of the joint temporal intensity, $\tilde{J}_I(t_s,t_i)$.  Biphoton states exhibiting entanglement can violate this inequality by an amount depending on its strength.  This strategy maps to the time-frequency domain tools which are commonly employed for quantifying entanglement in two-mode squeezed light. This demands accessing directly the temporal degree of freedom which can be done by time gating through a nonlinear optical process, as customary in time-resolved ultrafast pulse modulation. 

\begin{figure*}[t!]
\includegraphics[width= \textwidth]{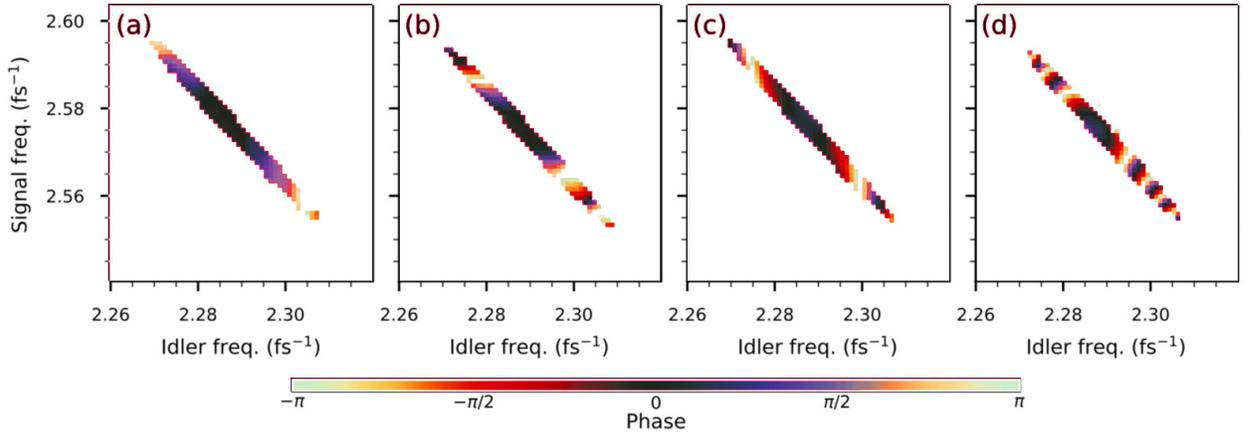}
\caption{{\it JSP reconstruction.} From left to right: reconstructed joint spectral phase for the case of unchirped photons (a),with the addition of a positive dispersion on the signal (b), in the case of a negative dispersion applied on the idler (c) and in the case of negative dispersion on both the photons (d). Figure adapted from Ref.\cite{Resch2019}.}
\label{figfasiresch}
\end{figure*}

Direct entanglement quantification via Eq. \eqref{prodotto} has been performed by K. J. Resh's group in Ref. \cite{John2018}. They consider a two-photon state generated by Type-I SPDC through a 2 mm  bismuth-borate (BiBO) crystal. The setup is engineered so that each of the two photons can undergo either a frequency or a time measurement. The pump is now a pulse of temporal width 120 fs, making it feasible to measure frequency correlations directly with a spectrometer. The frequency measurement is performed by means of a  grating-based monochromator with resolution of 0.1 nm. The gating for the time measurement is implemented with sum-frequency generation with a pick-off from the same laser used for SPDC; the temporal width of each photon in the pairs is sufficiently large for the pump to provide sufficient temporal resolution. The setup is sketched in Fig. \ref{figjohn}.

Data are recorded as coincidences counts either from the two spectrometers or the two temporal gating apparata.
By binning these into histograms according to $\omega_s+\omega_i$ and $t_s-t_i$ it is possible to evaluate the violation of Eq.~\eqref{prodotto} as an entanglement criterion.  The envelopes can be fitted with a Gaussian profile of widths  $\Delta(\omega_s+\omega_i)=(1.429\pm0.005)\mbox{ ps}^{-1}$ and $\Delta(t_s-t_i)=(0.203\pm0.005)\mbox{ ps}$ resulting in a violation given by  $\Delta(\omega_s+\omega_i)\Delta(t_s-t_i)=0.290\pm0.007$.  The recorded histograms show a time correlation between the two photons produced by SPDC, amounting to $0.987\pm0.004$, while they exhibit a strong anticorrelation in energy, equal to $-0.9951\pm0.0001$. Cross-correlation between time and frequency of signal and idler can also be recorded to monitor the presence of dispersion. These amount to $0.111 \pm 0.008$  and $-0.106 \pm 0.008$.

Second order spectral phases can be applied on both photons, as a way to manipulate the JSA. This is done by introducing grating compressors before the time gating. In fact, frequency measurements would be insensitive to such phases. The state of each individual photon, being it mixed, will also  be unaffected. The temporal correlations instead may be modified by the presence of such phases. These will result in broadening of the temporal width.  However, it is possible to achieve nonlocal dispersion cancellation by applying opposite phases to the two photons. Under such circumstances we expect time correlation properties to remain unaffected as well. 
A measurement of the temporal width in the presence of second order phases $(0.0373 \pm 0.0015)\,ps^2$ on the signal  and $(-0.0359 \pm 0.0014)\,p^2 $ lead to $\Delta(t_1-t_2)=0.245\pm0.004$ to be compared to the measurement without compressors $\Delta(t_1-t_2)=0.235\pm0.005$. This same technique has been proven useful for characterizing entanglement in quantum interference with subpicosecond delays \cite{Resch2018}.

\section{Reconstruction of the Joint Spectral Amplitude}

The methods described above are limited in that details of the two photon wavefunction cannot be appreciated. This has motivated investigations into the full characterization of the JSA.  Several of this rely on ideas borrowed from ultrafast techniques. These ideas lead to a different interpretation of the data collected in the experiment described in the previous section, which has been reported in Ref. \cite{Resch2019}. Indeed, the four measured traces $I$ in the time and frequency domains are interlinked by relations in the form

\begin{equation}
\begin{aligned}
&I(\omega_s,\omega_i)=\vert J_A(\omega_s,\omega_i)\vert^2\\
&I(\omega_s,t_i)=\left\vert\int d\omega_i \,J_A(\omega_s,\omega_i)e^{-i\omega_i t_i}\right\vert^2,
\end{aligned}
\end{equation}
with similar expressions holding for the other three cases.  It can be recognized that these bear the same structure as the spectrogram in Eq.~\eqref{spectrogram}, hence they can be processed in an analogous fashion, starting from a guess on the JSA.  

\begin{figure*}[t!]
\includegraphics[width=\textwidth]{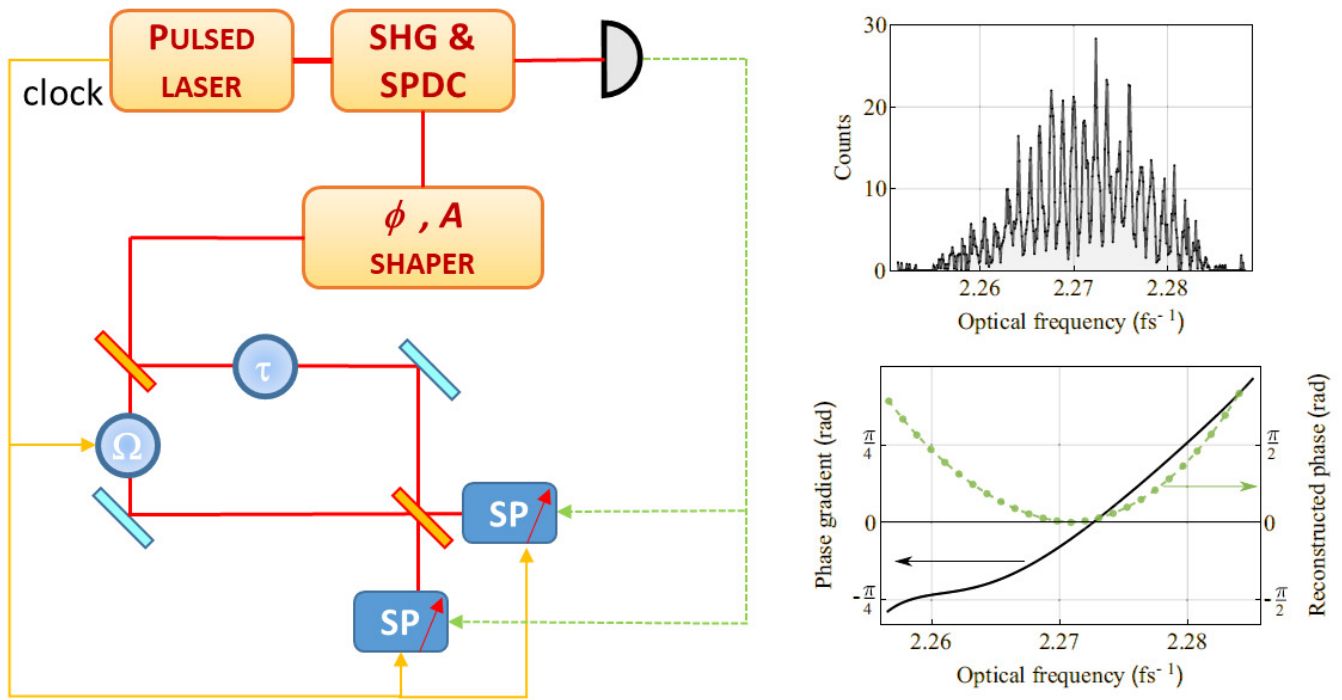}
\caption{{\it Single photon spectral shearing interferometry.} Left: experimental setup. Upper right: measured interferogram. Lower right: retrieved phase and gradient. Figure adapted from Ref. \cite{Brian1}.}
\label{brian}
\end{figure*}

The main difference with respect to the FROG iterative procedure in Fig. \ref{schemafrog} is that there is no need to operate the convolution and the deconvolution step to isolate the contribution of the gate. The remaining steps are iterated four times by Fourier Transforming one dimension at the time, and inferring constraints from the measured traces at each step, in the following order:
\begin{equation}
I(\omega_s,\omega_i) \xRightarrow{iFFT} I(\omega_s, t_i) \xRightarrow{iFFT} I(t_s,t_i) \xRightarrow{FFT} I(t_s,\omega_i),
\end{equation} 
where the transforms are taken on the amplitudes. We remark that differently from the standard FROG  traces, the field to be reconstructed is two dimensional, hence the two domains refer to distinct photons. 
Taking into account constraints from all four measured traces removes the time-reversal ambiguity in the reconstruction.  In Fig. \ref{figfasiresch} we report the biphoton phase reconstruction for different second order phases added to the two photons observed in Ref.~\cite{Resch2019}.  

Extending this iterative technique to more structured cases might be affected by the temporal resolution. For this purpose, it may become convenient to take a direct approach to measuring the spectral phase by applying SPIDER. To this aim one of the most challenging aspects of transferring ultrafast techniques to the quantum domain is that they all make use of  nonlinear phenomena: in spectrographics techniques they are exploited for temporal gating, in interferometric techniques they are necessary for providing the frequency shear, through upconversion at different times with a chirped ancillary beam. Due to the strong intensity limitations, straightforward application to the quantum regime is prohibitive. This can be mitigated using an external gating, as in the previous example, or by adopting alternative solutions to replace the nonlinear interaction. 

The quantum SPIDER approach, described in Ref \cite{Brian1,Brian1bis} for the reconstruction of heralded single photons, is a case in point. In these implementations the shear is applied by adding a linear temporal phase to the pulse by means of an electro-optical modulator (EOM) \cite{laura}:  due to the Fourier transform relations, the effect in the frequency domain of a linear temporal phase is indeed that of a frequency shear, in the same way that a linear spectral phase corresponds to a time delay. The mode at the output of the EOM will read: 

\begin{equation}
    \varepsilon_{out}(t)=\varepsilon_{in}(t)\cdot e^{\phi(t)}= \varepsilon_{in}(t)\cdot e^{-i\pi\frac{  V_{max}}{V_{\pi}}\sin{(2\pi f t)}},
\end{equation}
where $V_{\pi}$ is the voltage required to achieve a $\pi$ phase shift, $f=1/T_{RF}$ is the frequency of the RF driving field, and $V_{max}$ is the maximum voltage which can be applied to the EOM. For the temporal phase $\phi(t)$ to be linear, the temporal support of the photon must be shorter than the period of the RF field modulation $T_{RF}$, so that the added phase becomes

\begin{equation}
\phi(t) \simeq \frac{2\pi^2 f\,  V_{max}}{V_{\pi}}t = \Omega t
\end{equation}
where $\Omega$ is the frequency shear that will be imparted on the pulse. In spectral shearing interferometry the choice of the shear is of fundamental importance as due to the retrieval procedure it will result in the sampling rate of the reconstructed field. As such, on one hand it has to be small enough to comply to the Whittaker-Shannon sampling limit\cite{shannon,whittaker}: for a pulse with temporal support $T_{\varepsilon}$, the spectrum needs to be sampled at a maximum angular frequency given by $\Omega\leq 2\pi/T_{\varepsilon}$ which thus imposes an upper bound on the shear. On the other hand, the shear needs not being too small to compromise the signal-to-noise ratio for the retrieved phase difference $\Theta(\omega)$ which can then become too sensitive to shot-to-shot instabilities; as a rule of thumb, a shear that is a few percents of the full spectral bandwidth is desirable. These considerations provide bounds both on the maximum pulse duration and on the attainable amount of shear achievable via electro-optical modulation, identifying an operational window for the feasibility of the reconstruction.  A typical EOM shear is of the order of hundreds of GHz \cite{Wong1994, Dorrer2003}, so this technique is well suited to characterize photons with temporal bandwidths down to hundreds of fs.

A sketched version of the experimental scheme adopted by B.Smith's group is reported in Fig. \ref{brian} (a). They rely on SPDC from a BiBO crystal, and build am EOM-shearing interferometer for one of the two photons, using the second photon for heralding as they record the frequency-dependent coincidence counts. In order to test the technique they employ a spatial light modulator (SLM) for pulse shaping, obtaining the reconstruction for different phase orders imparted on the pulse with the SLM. An example of a measured interferogram is shown in Fig. \ref{brian} (b). The phase retrieval algorithm follows the same steps as the SPIDER algorithm detailed earlier. Fig. \ref{brian} (c) shows one of the reconstructed phases. 
This first implementation only considered the marginal of the biphoton wavefunction, as it was not interest in conveying information on the correlations between the two photons. The technique has recently been applied in Ref. \cite{Brian2} to the reconstruction of the full biphoton wavefunction.

Despite the use of alternatives to nonlinear processes helps tempering the low-signal condition, the rate of recorded coincidences per frequency-bin still struggles to exceed a few Hz. As mentioned earlier, an additional element which can be employed to improve the SNR in the reconstruction is the use of redundant measurements. This has been proven classically, where multi-shear techniques have outperformed single-shear ones, as SEA-CAR SPIDER implementations and, more recently, MICE. 
However, adapting these to the quantum domain is not straightforward and requires to identify the appropriate measurement strategy to do so, as well as to account for the tighter instrumental resolution restrictions. This has been tackled in Ref. \cite{Frice1,Frice2}, where a quantum adaptation of the MICE technique is proposed. 

The experimental scheme relies on a modified Franson interferometer, which outputs a polarization-entangled state which provides the correct interference term for the MICE reconstruction, albeit with some modification: contrary to the classical case, the field to be retrieved is two dimensional; this fits well within the capabilities of MICE, however it requires a variable shear to be applied along each dimension. As in the previous example, the shears can be exerted through electro-optical modulation. Simulations of phase reconstruction underline a strong robustness to noise, with the phase correctly retrieved even with a maximum of 5 coincidences per frequency bin. 

\begin{figure}[h!]
\includegraphics[width= \columnwidth]{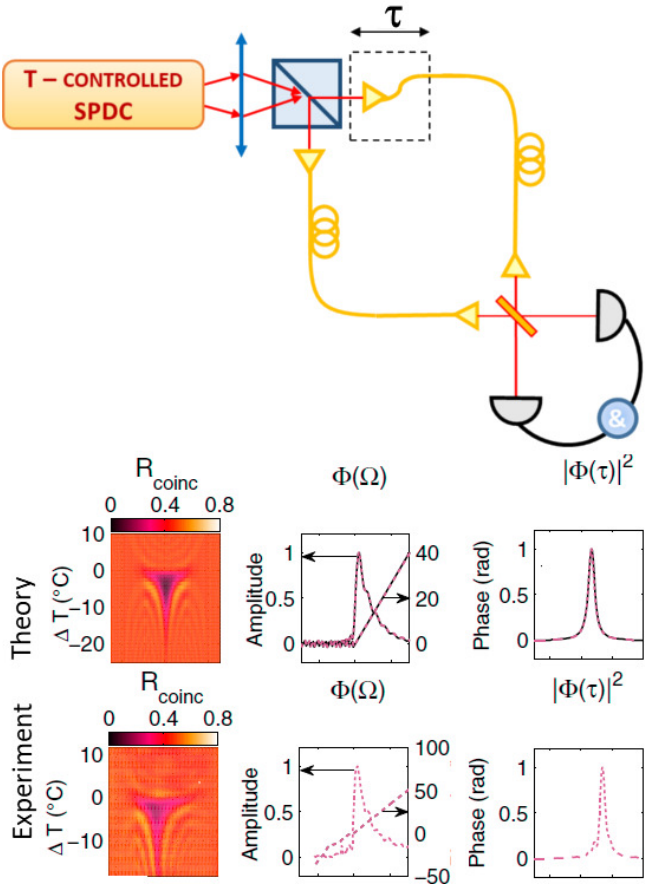}
\caption{Upper) experimental setup: the two photons produced by a temperature controlled SPDC source are collimated and then separated by a polarising beam splitter in the two arms of an interferometer set for an HOM scheme, where a relative delay $\tau$ is imparted. Coincidence counts are recorded. Lower) An example of the reconstruction of the wavefunction and time delay distribution. Adapted from~\cite{MolinaTerriza15}}
\label{figmolina}
\end{figure}

All measurements described so far implement interferometric and spectrographic techniques directly inherited from the classical world. One might seek to apply reconstruction methods exploiting quantum features in Hong-Ou-Mandel (HOM) interferometry\cite{HOM}, as this is a commonplace technique when working with few photons \cite{ben,pan}. In a HOM experiment, two photons arrive at the surface of a symmetric beamsplitter with a relative delay $\tau$, and coincidences are measured between the two output modes. If the JSA is symmetric, at zero delay no coincidences will be registered \footnote{This instance includes and generalizes the statement that photons must be completely indistinguishable to get such perfect cancellation: this is actually the case only when photons are in a separable state.}.

The shape of the curve of the coincidences as a function of the delay, often referred to as the HOM profile, will depend on the JSA, and might contain some relevant information \cite{Fedrizzi2009a,Eckstein2008,Wang2006}. For instance if the coincidence rate exceeds the value for random independent splitting, some entanglement is bound to be present. However, the actual information content is rather low and general considerations on the JSI, let alone on the JSA are hard to be inferred, even when complemented by extra observations.  Nevertheless this is true because one observes only one set of conditions. By accumulating different HOM profiles taken at different phase matching conditions, it is possible to collect enough information for a successful reconstruction of the JSA. 

In Ref.\cite{MolinaTerriza15} the group of G. Molina-Terriza has reconstructed the JSA of a CW 404.25 nm pumped SPDC source from a Type-II periodically poled potassium titanyl phosphate crystal. The narrow width along the $\omega_s+\omega_i$ direction allows to treat this problem as one-dimensional, focusing only on the $\omega_-= \omega_s-\omega_i$ direction. In this approximation the HOM profile as a function of the delay $\tau$: 

\begin{equation}
R \propto \int d\omega_- \Phi(\omega_-; \delta\omega_0,\omega_p)\Phi^*(-\omega_-; \delta\omega_0,\omega_p)e^{i\omega_-\tau}.
\end{equation}

In this formula the phase matching is set for the central frequencies of the signal and idler photons to differ by an amount $\delta\omega_0$. This profile is collected as a function of $\tau$ for different values of $\delta\omega_0$. This scheme  then takes inspiration from FROG in the delay-scanning, as well as SPIDER, by applying multiple shears between the two photons. In Fig. \ref{figmolina} we show the data collected in Ref \cite{MolinaTerriza15} along with the reconstructed one-dimensional JSA.

\begin{figure}[b!]
\includegraphics[width=\columnwidth]{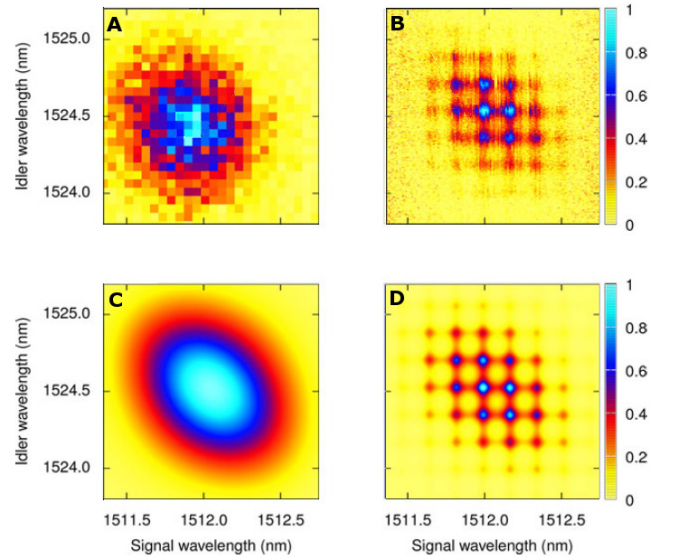}
\caption{Experimental reconstruction of JSI using conventional frequency coincidence counting (panel A) and the SET approach (panel B). Panel C is the convolution of a Gaussian (FWHM = 224 pm) with the calculated JSI for the SPDC process shown in panel D. Adapted from Ref. \cite{andreas2014}.}
\label{figandreas}
\end{figure}

We observe that tomographic methods have been developed for investigating single-photon spectral wavefunctions. These are based on interference of single photons with a local oscillator, and detection based on either photon counting\cite{frankowski}, or homodyne detection~\cite{bellini1}. These have not yet been applied to assess spectral correlations.

\section{Stimulated emission tomography}

Overall, all methods based on the direct characterization of the JSA described up to now try to overcome the low-signal restrictions associated to the very same fact of investigating light in the quantum regime. Although the SPDC process is quintessentially quantum, it can be interpreted as a difference frequency generation (DFG), a process well understood in classical terms, between the pump and the vacuum state \cite{Helt2012}; as a result, the two nonlinear phenomena share the same phasematching function.  Within this framework, under the same pumping conditions, it is possible to infer information on the spectral correlations generated through the SPDC by performing stimulated emission tomography (SET) \cite{liscidini2013}. This transposes to the classical regime the problem of assessing frequency correlations, by stimulating DFG of a classical, monochromatic signal, and looking at the spectral content of the generated idler. The advantage also allows to improve the spectral resolution, by relying on conventional spectrometers. 

This technique has known several implementations \cite{Rozema2015,Fang2014,setmario}; here we review the one by S. Ducci's group in Ref.\cite{andreas2014} as an illustrative example, where SET has been compared to standard JSI retrieval via coincidence counting spectrometers. There, they inject the same CW pump beam in a Aluminium gallium arsenide (AlGaAs) waveguide. In the SPDC scheme, signal and idler photons are then collected and sent through two fiber single photon spectrometers \cite{xstine2009}, as described earlier. In the DFG scheme, they seed the signal mode with a second CW laser. With this geometry it is possible to record with a conventional spectrometer the "slices" of the idler spectrum at a given signal frequency given by that of the CW seed,  which will eventually form the complete JSI as the signal seed is frequency-scanned. The reconstruction of the JSI with the two techniques is shown in Fig. \ref{figandreas}. The comparison is extremely favourable for the SET approach, also due to the improvements in the SNR. This advantage is by all means not peculiar to this one implementation but it is rather a common feature intrinsic to the use of classical beams. 

\begin{figure}[t!]
\includegraphics[width=\columnwidth]{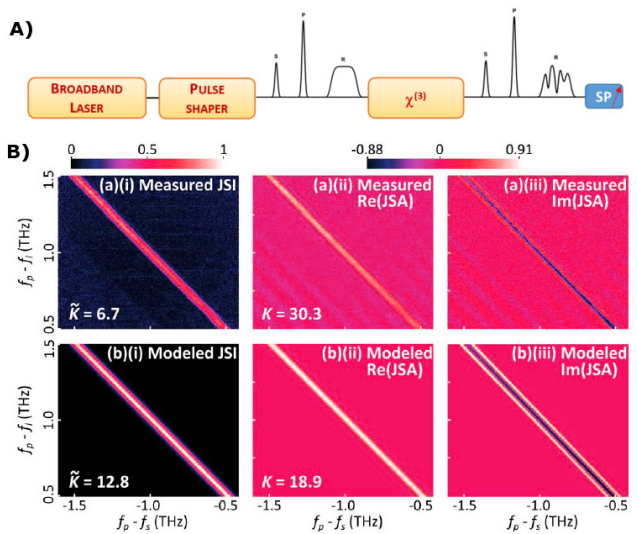}
\caption{{\it FWM full JSA retrieval} a) experimental setup for the measurement of the JSA: a broadband laser is shaped to produce three pulses: a pump (P), a seed (S) and a reference (R) pulse. b) JSI and both real and imaginary parts of JSA. Adapted from \cite{Jizan2016}.}
\label{figegg}
\end{figure}

The same technique can also be applied to the retrieval of the complex JSA, when complemented with spectral interferometry. This concept has been explored experimentally by the group of B.J. Eggleton in Ref. \cite{Jizan2016}. There, as described in Fig. \ref{figegg} (a) they consider the JSA produced by spontaneous four-wave mixing in a nonlinear waveguide (either a Silicon nanowire or a Hollow-core nonlinear fiber). Their collinear scheme is based on the employment of a liquid-crystal-on-silicon dynamically tunable filter (LCoSWS) to shape a broadband laser centered at 1555 nm with a 30nm bandwidth. The LCoSWS cuts it in three distinct pulses: the pump (a Gaussian centered at 1553) , the seed (frequency scanned with a spectral width of 10GHz which determines the signal spectral resolution) , and a reference (with a bandwidth of over 1THz), and at the same time serves to control the linear (delay) and quadratic (dispersion) spectral phase of the three pulses. 
The protocol consists in injecting seed pulses with different phase shifts ($\theta=0,\pi/2,\pi,3\pi/2$) obtaining four different interferograms as the result of the superposition of the generated idler and the reference pulse. By combining these four interferograms it is then possible to retrieve the real and complex part of the JSA, as shown in Fig. \ref{figegg} (b). 

\section{Conclusions}

Characterization of time-frequency properties is key to future development in quantum photonics. 
In the classical domain, time-frequency characterization has been a longstanding problem in ultrafast metrology, and the know-how accumulated in the last few decades can serve as an important background to underpin novel approaches in the quantum domain. 
Here we have provided a short review of the different techniques which can be employed to quantify frequency correlations and to characterize time-frequency states. The implementations considered are either techniques stemming from purely quantum considerations, or adaptation of classical ultrafast metrology approaches, which have been modified to enable the complete characterization of the biphoton JSA.


%
%

%

\begin{acknowledgments}
\end{acknowledgments}

\bibliography{frequencytime.bib}

\end{document}